\documentstyle[12pt,psfig,draft,lscape,array,amssymb,longtable]{nature-ejb}

\def\simlt{\mathrel{\hbox{\rlap{\hbox{\lower4pt\hbox{$\sim$}}}\hbox{$<$}}}}
\def\simgt{\mathrel{\hbox{\rlap{\hbox{\lower4pt\hbox{$\sim$}}}\hbox{$>$}}}}
\def\ale{\mathrel{\hbox{\rlap{\hbox{\lower4pt\hbox{$\sim$}}}\hbox{$<$}}}}
\def\age{\mathrel{\hbox{\rlap{\hbox{\lower4pt\hbox{$\sim$}}}\hbox{$>$}}}}

\begin{document}
\title{\Large An unusually brilliant transient in the galaxy Messier\,85}
\vspace{-0.1in}
\author{\small S. R. Kulkarni\affiliation[1]
	{Caltech Optical Observatories 105-24, 
	California Institute of Technology, Pasadena, CA\,91125,
	USA\vspace{0.1in}},
	E.~O.~Ofek\affiliationmark[1],
        A.~Rau\affiliationmark[1],
	S.~B.~Cenko\affiliation[2]
	{Space Radiation Laboratory 220-47, 
	California Institute of Technology, CA\,91125, USA\vspace{0.1in}},
        A.~M.~Soderberg\affiliationmark[1],
	D.~B.~Fox\affiliation[3]
	{Department of Astronomy, 
        Pennsylvania State University, 
	State College, PA\,16802, USA\vspace{0.1in}},
        A.~Gal-Yam\affiliationmark[1],
        P.~L.~Capak\affiliationmark[1],
        D.~S.~Moon\affiliationmark[2],
	W.~Li\affiliation[4]
	{Astronomy Department, 601 Campbell Hall, University of California,
	Berkeley, CA\,94720, USA \vspace{0.1in}},
	A.~V.~Filippenko\affiliationmark[4],
        E.~Egami\affiliation[5]
	{Steward Observatory, 933 N Cherry Avenue, University of Arizona, Tucson AZ 85721, USA\vspace{0.1in}},
	J.~Kartaltepe\affiliation[6]
	{Institute of Astronomy, University of Hawaii, 
	2680 Woodlawn Drive, Honolulu, HI 96822, USA}
        \&\
	D.~B.~Sanders\affiliationmark[6]
}
		
\dates{2006 May 14}{}
\headertitle{Brilliant transient in Messier\,85}
\mainauthor{Kulkarni et al.}

\summary{ 
Historically, variable and transient sources have both surprised
astronomers and provided new views of the heavens.  Here we report
the discovery of an optical transient  in the outskirts of the
lenticular galaxy Messier 85 in the Virgo Cluster.  With a peak
absolute $R$ magnitude of $-12\,$ this event is distinctly brighter
than novae, but fainter than type Ia supernovae (expected from
a population of old stars in lenticular galaxies).  Archival images
of the field do not show a luminous star at that position  with an
upper limit of $g\sim-$4.1, so it is unlikely to be a giant eruption
from a luminous blue variable star.  Over a two month period the
transient emitted radiation energy of almost 10$^{47}$\,erg and
subsequently faded in the opical sky. It is similar to, but more
luminous at peak by a factor of 6 than, an enigmatic transient in
the galaxy M31.  A possible origin of M85\,OT2006-1 is a stellar
merger. If so, searches for similar events in nearby galaxies 
will not only allow to study the physics of hyper-Eddington sources,
but also probe an important phase in the evolution of stellar
binary systems.  
}

\maketitle

\begin{figure}[ht]
\centerline{\psfig{file=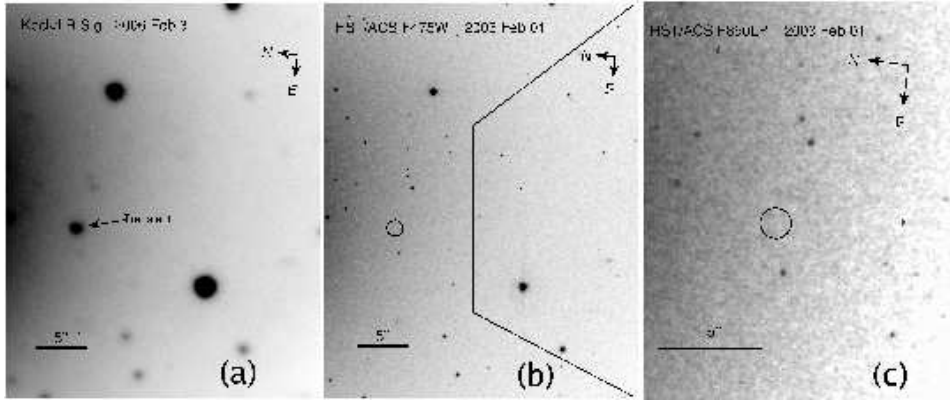,width=5in,angle=0}}
\caption[]
{{\small Optical images of the field around M85\,OT2006-1 obtained
at two epochs. Data were 
obtained with the Low-Resolution Imager and Spectrograph 
(LRIS;\cite{occ+95}) at Keck 
(a) on Feb 3, 2006, and the Advanced Camera for 
Surveys aboard the Hubble Space Telescope (b - F475W filter; c - 
F850LP filter) on Feb 1, 2003. The event is located about 30" from the 
center of M85 at $\alpha$=${\rm 12^h 25^m 23.82^s}$ and $\delta$=$18^\circ 
10^\prime56.2^{\prime\prime}$ (J2000). 
After registering the Keck image to the HST image (rms of the
transformation was 40 mas) we were able to place the following
limits for
a pre-cursor object (progenitor star): $26.8\,$mag in the F475W filter 
(exposure 750\,s) and $24.7\,$mag in the F850LP filter (exposure time 
1150\,s). These limits exclude an LBV\cite{hd94} 
origin (for which $M_{\rm V}\sim -8\,$mag).  Furthermore, we find no evidence for 
young stars (supergiants, clusters and HII regions).  An analysis of 
Spitzer Space Telescope Infared Array Camera data obtained 
on Dec 21, 2004,  result in 3$\sigma$ upper limits of 25, 30, 60 and 75\,$\mu$Jy at 3.6, 
4.5, 5.8 and 8.0\,$\mu$m, respectively.
{\it LOSS} observed M85 two hundred and twenty
times over 2000--2006.  We found no transient at the position of
M85\,OT2006-1 to (roughly $R$-band) magnitudes ranging from 20 
to 21. 
  No X-ray emission was detected 
in a {\it Chandra} X-ray Observatory observation\cite{ssi+03} obtained in 
June, 2002, with a flux upper limit of $2.7\times10^{-4}\,$cnt~s$^{-1}$ in 
the 0.3--10~keV band.}}
\label{fig:M85_HSTKeck} 
\end{figure}

On January 7, 2006, the Lick Observatory Supernova Search {\it(LOSS)}
team, in their daily circular, reported a source with apparent
unfiltered magnitude of $\sim19.3$ projected 2.3\,kpc from the
center of the lenticular (S0) galaxy Messier\,85 (M85; also known
as NGC~4382)\footnotemark\footnotetext{see
\texttt{http://nedwww.ipac.caltech.edu}}, a member of the Virgo
cluster of galaxies; see Figure~\ref{fig:M85_HSTKeck}.  
For reasons stated below we believe that the transient lies within
M85 and thus we name the source as 
M85\,Optical Transient 2006-1 (or M85\,OT2006-1 for
short). Fortuitously this field was observed by the {\it Hubble Space
Telescope} (HST) three years earlier.  From the archival HST data we 
derive a pre-explosion
limiting magnitude of 
F475W=$-4.3$.

\begin{figure}[ht]
\centerline{\psfig{file=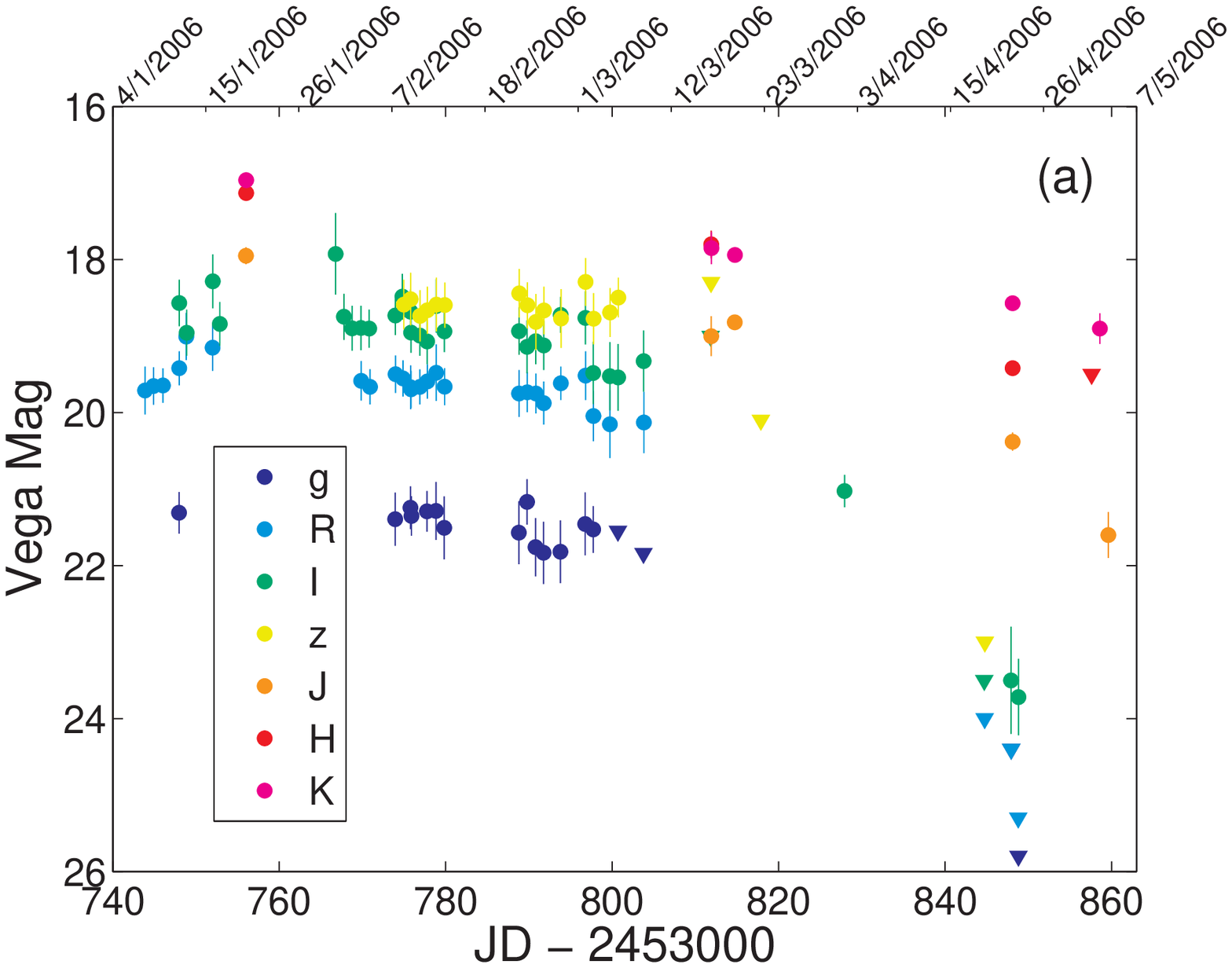,width=5in,angle=0}}
\centerline{\psfig{file=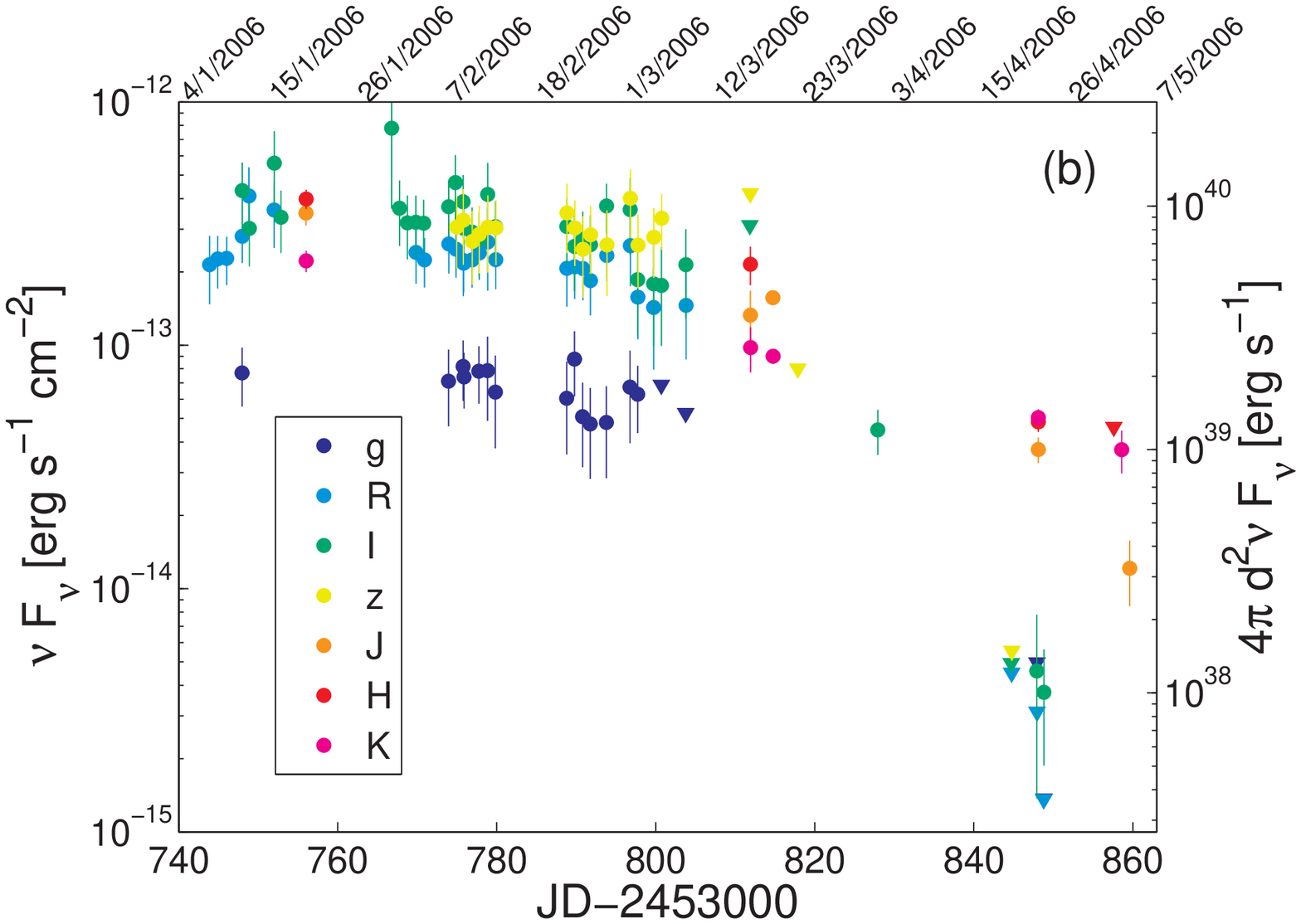,width=5.5in,angle=0}}
\caption[]
{{\small Temporal evolution of
 M85\,OT2006-1. (a) observed light curve uncorrected for Galactic foreground extinction and (b) $\nu f_\nu$ including foreground extinction correction of $A_{V}=0.08$\cite{sfd98}. 
Data for the plots are given in Photometry Table of the electronic
supplement and come from the following sources: Palomar 60-inch
(P60; $gRIz$), the Large Format Camera (LFC) on the Palomar Hale 200-inch
(P200; $zRI$), the Widefield Infrared Camera (WIRC) on P200 ($JHK$),
LRIS on the Keck-I 10-m telescope ($gRI$), Persson's Auxilary Nasmyth
Infrared Camera (PANIC) on the Magellan 6.5-m Baade telescope ($YJK$),
Near Infared Echelle Spectrograph (NIRSPEC) on the Keck-II 10-m
telescope ($JHK$) and the Wide Field Infrared Camera (WFCAM) on the
3.8-m United Kingdom Infrared Telescope (UKIRT; $JHK$).}}
\label{fig:LC} 
\end{figure}

On UT 2006~January~8 we initiated an optical photometric campaign
with the automated Palomar 60-inch telescope (see Figure~\ref{fig:LC}
and Table in the electronic supplementary material).
The light curve with a plateau of 70 days 
is unlike that of type Ia supernova. The plateau duration is also too short
for an outburst from a Luminous Blue Variable (LBV;  $\eta$ Carina). 

\begin{figure}[htb]
\centerline{\psfig{file=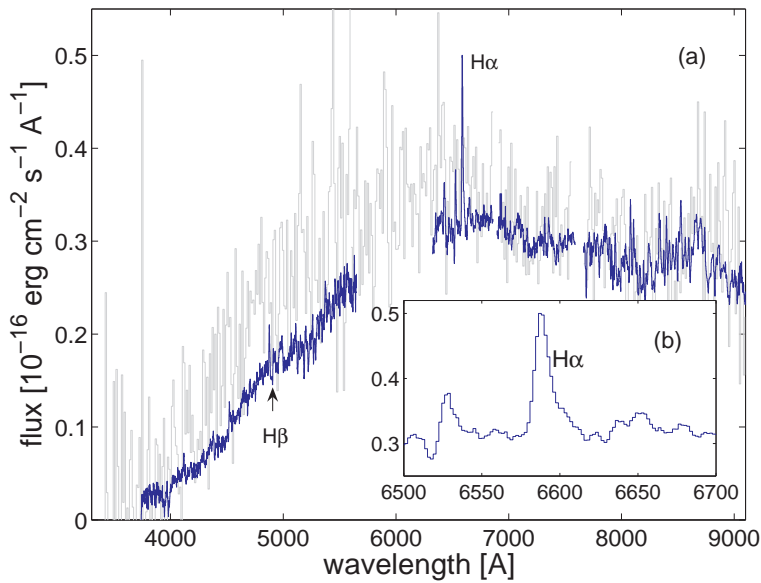,width=4in}}
\caption[]
{{\small Optical spectra of M85\,OT2006-1. Data were obtained with the Double Beam Spectrograph\cite{oke81} (DBSP) at the Palomar Hale 
200-inch telescope (grey line, 1800\,s integration, Jan 8.53 UT, 2006) and Keck/LRIS 
(blue line, 3000\,s, Feb 24.59 UT, 2006). Not strong emission or absorption 
features are seen in the (native) DBSP spectrum. Specifically we place a limit of $6\times 10^{-16}$\,erg 
s$^{-1}$\,cm$^{-2}$
for an emission line in the vicinity of H$\alpha$. 
In the LRIS red channel spectrum the brightest emission feature is
at $\lambda=6587$\,\AA\     (flux of $(3.2\pm0.2) \times
10^{-16}\,$erg\,s$^{-1}$\,cm$^{-2}$) which we  identify with
redshifted H$\alpha$. The velocity of the line center is $1020\pm
150$\,km\,s$^{-1}$ (see inset). On the blue side, the strongest
feature is at $\lambda=4875$\,\AA\, corresponding to 
redshifted ($700\pm100\,$km\,s$^{-1}$) H$\beta$ (flux of $(0.9\pm0.1) \times
10^{-16}\,$erg\,s$^{-1}$\,cm$^{-2}$) .   The  equivalent widths  are
$10\pm1$\,\AA\ (H$\alpha$) and  $5\pm1$\,\AA\ (H$\beta$).  The full width
at half maximum (FWHM) of the H$\alpha$ line, after accounting for
the instrumental FWHM, is $350\pm 140$\,km\,s$^{-1}$.

In  addition we  detect the following emission  lines
(central wavelengths, typical uncertainty of 1\,\AA;  and
fluxes, unit of $10^{-16}\,$erg\,s$^{-1}$\,cm$^{-2}$):
4115\,\AA\ ($0.3\pm 0.1$), 6428\,\AA\  ($0.9\pm 0.1$), 6527\,\AA\
($1.5\pm 0.4$), 8079\,\AA\ ($0.8\pm 0.1$) and 8106 ($0.7\pm 0.1$).

Further LRIS  spectra were obtained on  UT 2006 February 3 and 23
(not shown here).   The February 3rd  LRIS  spectrum did not include
the H$\alpha$ wavelength.  For this spectrum, using  a sliding
10\,\AA\  window we were able  to set  a 3-$\sigma$ upper limit  of
$\simlt  6\times 10^{-18}\,$erg cm$^{-2}$  s$^{-1}$ in the vicinity
of H$\beta$.
}}
\label{fig:SpectrumDBSPLRIS} 
\end{figure}

We began a program of spectroscopic observations with the Palomar
Hale and the Keck I  telescopes (Figure~\ref{fig:SpectrumDBSPLRIS}).
The Palomar spectrum obtained on UT 2006 January 8 did not contain
any strong emission feature; the spectrum could be adequately
described by  a black body spectrum with effective temperature of
approximately $T_{\rm eff}\sim 4600\,$K. Likewise the UT 2006
February 3 Keck spectrum  was also featureless but unfortunately
did not cover H$\alpha$.  The UT 2006 February 23 and 24 Keck spectra
showed a similar continuum but a number of emission lines were
readily detected (Figure~\ref{fig:SpectrumDBSPLRIS}).  Since the
latter spectra were the deepest it is likely that the lines were
seen due to better sensitivity.

We associate the strongest line at wavelength, $\lambda\simeq
6587$\,\AA\ and that at $\lambda\simeq 4874$\,\AA\ with H$\alpha$
and H$\beta$, respectively.  Accepting this identification, the
mean heliocentric (peak) velocity of the pair is $880\pm 130$
km\,s$^{-1}$.  We were unable to conclusively identify the remaining
lines but do note that the spectra of many hypergiants contain a
number of unidentified\cite{hd94,hjp+06} emission lines.

The systemic velocity\cite{slh+00} of M85 is $729\pm2$\,km\,s$^{-1}$
and the velocity dispersion\cite{ecp+04} in the vicinity of the
optical transient is 200\,km\,s$^{-1}$.  The peak velocity of the
Balmer lines is thus consistent with M85\,OT2006-1 being located
in M85.  
Accepting that M85 is the host galaxy (for which we adopt
a distance of 15\,Mpc, the standard distance to the Virgo
cluster\cite{fmg+01}) the absolute $R$-band magnitude of M85\,OT2006-1
is $-12\,$mag.  This peak flux is  ten times brighter than the
brightest nova but (at least) ten times less luminous than supernova
of the Type Ia (the sort expected in a lenticular galaxy).
The
narrow line width of the H$\alpha$ line,
$\sim 350\pm 140\,$km\,s$^{-1}$ (see Figure~\ref{fig:SpectrumDBSPLRIS}),
argue independently against both a  nova and a supernova (including
of type II) origin.

The Galactic foreground extinction towards M85 is negligible, $A_{\rm
R}=0.08$\cite{sfd98}. The source intrinsic attenuation  can be
derived by comparing the  observed ratio of the emission lines
fluxes of H$\alpha$ [$(3.2\pm0.2) \times
10^{-16}\,$erg\,s$^{-1}$\,cm$^{-2}$] and H$\beta$ [$(0.9\pm0.1)
\times 10^{-16}\,$erg\,s$^{-1}$\,cm$^{-2}$] and the theoretical
value of 3.05\footnote{Case B recombination\cite{o89}, low-density limit,
$T=5000\,$K.}. We estimate $E(B-V) = 0.14_{-0.14}^{+0.17}$
which corresponds to an $R$-band extinction of $0.40_{-0.40}^{+0.48}$\,mag.
This is to low to explain the unusual color and temperature of
M85\,OT2006-1 with a strongly absorbed nova, supernova or LBV.

We searched archival data from HSTe,
the Spitzer Space Telescope and the Chandra X-ray Observatory  
with the view of constraining the progenitor.
There is no evidence for
a bright progenitor nor do we see tracers of massive star progenitors
(see Figure~\ref{fig:M85_HSTKeck}). This finding (and the shorter duration) rule out
that M85\,OT2006-1 is an LBV because 
LBVs are amongst\cite{hd94} the brightest stars, $M_{\rm V}<-8$.
Along these lines we note that M85 is a galaxy composed of old stars with a 
possible trace of
a spiral arm. We conclude that the M85\,OT2006-1 likely arises
from a population of stars with mass of few $M_\odot$ or smaller.

We now turn to the physical parameters of M85\,OT2006-1.
The bolometric luminosity flux (as traced by $4\pi d^2\nu f_\nu$;
here, $f_\nu$ is the spectral flux density at frequency $\nu$) of
M85\,OT2006-1 peaks at $L_p\sim2\times 10^{40}\,$erg\,s$^{-1}$.  Over
the first two months the total radiated energy is about $E_{\rm ph}
\sim 6\times 10^{46}\,$erg.  The inferred blackbody radius of the
object is substantial, $R=[L_p/(4\pi\sigma_B T_{\rm eff}^4)]^{1/2}\sim
17 (T_{\rm eff}/4600\,{\rm K})^{-2}\,$AU.

\begin{figure}[ht]
\centerline{\psfig{file=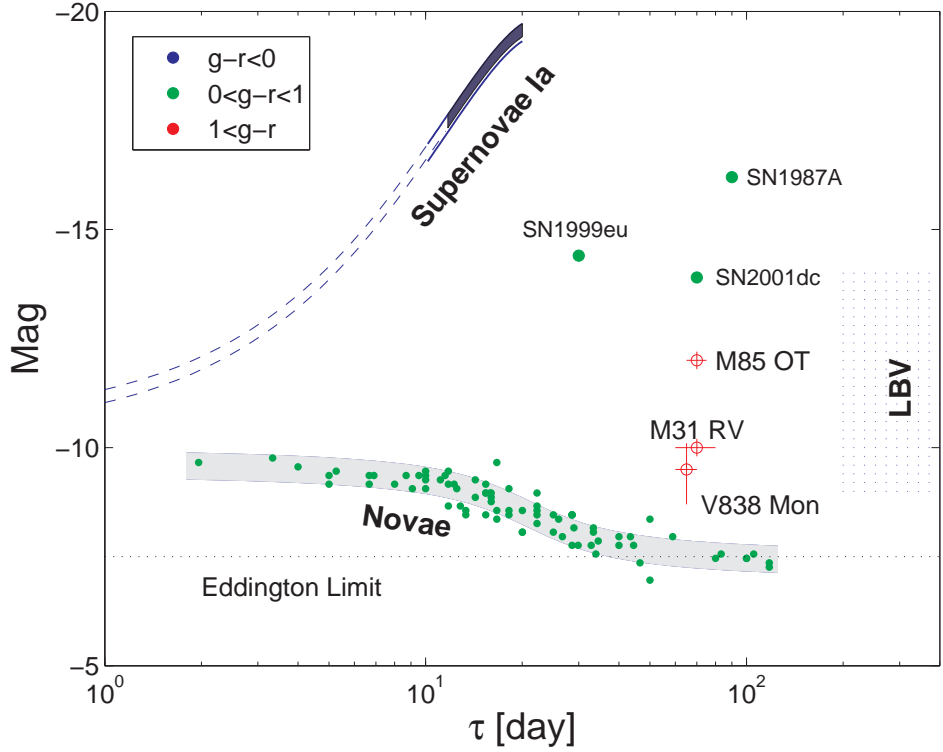,width=5in}}
\caption[]
{{\small Phase space of cosmic explosive and eruptive transients.  The
vertical axis is the peak brightness in the $R$-band and the
horizontal axis is the duration of the event ($\tau$).  Events are
represented by circles with the color at peak magnitude coded as
follows: blue ($g-r\simlt 0$), green ($1\simgt g-r\simgt 0$) and
red ($g-r\simgt 1$).  M85\,OT2006-1 and the M31\,RV\cite{rmp+89}
clearly stand out in this figure in the following respects: ({\it
i}) brighter than novae but ({\it ii}) less luminous than most
supernovae (especially of type Ia indicated with a 2$\sigma$
brightness band) and ({\it iii}) distinctly red color when compared
to sub-luminous core collapse supernovae (such as SN 1987A). Finally, 
the two events,
unlike LBVs and core collapse supernvoae,  do not arise in star-forming
regions. For any reasonable progenitor mass, both events exhibit
hyper-Eddington peak luminosities, similar to the sources
V838~Mon\cite{rm03,tylenda05}.  Furthermore, both sources also
characterized by low expansion velocity ($<1000\,$km\,s$^{-1}$) and
a strong redward evolution of the peak frequency. For these objects,
$\tau$, is the ``plateau" time scale.  For novae, $\tau$ is the
time scale in which the nova fades by two magnitudes, $t_{2}$.
Filled circles show the positions of 82 novae observed\cite{arp56,cdr+89}
in Messier\,31 (assuming\cite{dbk+04} $V-R=0.56$ at peak). The
brightest stars in our Galaxy are highly variable but these objects
(marked ``LBV'') are clearly distinguished by long variability
timescales and a high quiescent magnitude. The dashed line
($R=-7.5\,$mag) is the Eddington limit for a 1\,$M_\odot$ G-type
star.
}}
\label{fig:Mtau} 
\end{figure}

The closest analog to M85\,OT2006-1 is M31\,RV, a bright
event\cite{rmp+89} (serendipitously)  found in the bulge of Messier\,31 and still lacking
a satisfactory explanation.  The extra-ordinary brilliance of
M85\,OT2006-1 (Figure~\ref{fig:Mtau}) makes it doubly mysterious.
The Galactic transient V838~Mon\cite{bws+02},
while considerably  less luminous (see Figure~\ref{fig:Mtau}),
exhibit similar plateau light curves and redward evolution of the
broad-band spectrum.

The distinctive physical parameters
(relative to novae and supernovae; see Figure~\ref{fig:Mtau}) and
the potential connection to a fundamental stellar process (merger)\cite{tylenda05}
may warrant coining a name.  We suggest the simple
name {\it luminous red nova} with the adjectives highlighting 
the principal characteristics of M85\,OT2006-1.
Statistics (including especially the nature of the
host galaxies) and follow up studies would help astronomers unravel
the origin of these enigmatic transients and also study the physics
of hyper-Eddington sources.

\bibliographystyle{nature}
\bibliography{journals,m85}

\begin{thebibliography}{10}

\bibitem[{Humphreys} \& {Davidson}<1>]{hd94}
{Humphreys}, R.~M. \& {Davidson}, K. {The luminous blue variables:
  Astrophysical geysers}.
\newblock {\it Publ. Astr. Soc. Pacific} {\bf 106}, 1025--1051 October 1994.

\bibitem[{Humphreys} {\it et~al.}<2>]{hjp+06}
{Humphreys}, R.~M., {Jones}, T.~J., {Polomski}, E., {Koppelman}, M., {Helton},
  A., {McQuinn}, K., {Gehrz}, R.~D., {Woodward}, C.~E., {Wagner}, R.~M.,
  {Gordon}, K., {Hinz}, J.  \& {Willner}, S.~P. {M33's Variable A: A Hypergiant
  Star More Than 35 YEARS in Eruption}.
\newblock {\it Astron. J.} {\bf 131}, 2105--2113 April 2006.

\bibitem[{Smith} {\it et~al.}<3>]{slh+00}
{Smith}, R.~J., {Lucey}, J.~R., {Hudson}, M.~J., {Schlegel}, D.~J.  \&
  {Davies}, R.~L. {Streaming motions of galaxy clusters within 12000 km
  s$^{-1}$ - I. New spectroscopic data}.
\newblock {\it Mon. Not. R. astr. Soc.} {\bf 313}, 469--490 April 2000.

\bibitem[{Emsellem} {\it et~al.}<4>]{ecp+04}
{Emsellem}, E., {Cappellari}, M., {Peletier}, R.~F., {McDermid}, R.~M.,
  {Bacon}, R., {Bureau}, M., {Copin}, Y., {Davies}, R.~L., {Krajnovi{\'c}}, D.,
  {Kuntschner}, H., {Miller}, B.~W.  \& {de Zeeuw}, P.~T. {The SAURON project -
  III. Integral-field absorption-line kinematics of 48 elliptical and
  lenticular galaxies}.
\newblock {\it Mon. Not. R. astr. Soc.} {\bf 352}, 721--743 August 2004.

\bibitem[{Freedman} {\it et~al.}<5>]{fmg+01}
{Freedman}, W.~L., {Madore}, B.~F., {Gibson}, B.~K., {Ferrarese}, L., {Kelson},
  D.~D., {Sakai}, S., {Mould}, J.~R., {Kennicutt}, R.~C., {Ford}, H.~C.,
  {Graham}, J.~A., {Huchra}, J.~P., {Hughes}, S.~M.~G., {Illingworth}, G.~D.,
  {Macri}, L.~M.  \& {Stetson}, P.~B. {Final Results from the Hubble Space
  Telescope Key Project to Measure the Hubble Constant}.
\newblock {\it Astrophys. J.} {\bf 553}, 47--72 May 2001.

\bibitem[{Schlegel}, {Finkbeiner} \& {Davis}<6>]{sfd98}
{Schlegel}, D.~J., {Finkbeiner}, D.~P.  \& {Davis}, M. {Maps of Dust Infrared
  Emission for Use in Estimation of Reddening and Cosmic Microwave Background
  Radiation Foregrounds}.
\newblock {\it Astrophys. J.} {\bf 500}, 525--553 June 1998.

\bibitem[{Osterbrock}<7>]{o89}
{Osterbrock}, D.~E.
\newblock {\it {Astrophysics of gaseous nebulae and active galactic nuclei}}.
\newblock Research supported by the University of California, John Simon
  Guggenheim Memorial Foundation, University of Minnesota, et al.~Mill Valley,
  CA, University Science Books, 1989, 422 p. (1989).

\bibitem[{Rich} {\it et~al.}<8>]{rmp+89}
{Rich}, R.~M., {Mould}, J., {Picard}, A., {Frogel}, J.~A.  \& {Davies}, R.
  {Luminous M giants in the bulge of M31}.
\newblock {\it Astrophys. J.} {\bf 341}, L51--L54 June 1989.

\bibitem[{Brown} {\it et~al.}<9>]{bws+02}
{Brown}, N.~J., {Waagen}, E.~O., {Scovil}, C., {Nelson}, P., {Oksanen}, A.,
  {Solonen}, J.  \& {Price}, A. {Peculiar variable in Monoceros.}
\newblock {\it IAU Circ} {\bf 7785}, 1--1 January 2002.

\bibitem[{Tylenda}<10>]{tylenda05}
{Tylenda}, R. {Evolution of V838 Monocerotis during and after the 2002
  eruption}.
\newblock {\it Astr. Astrophys.} {\bf 436}, 1009--1020 June 2005.

\bibitem[{Oke} {\it et~al.}<11>]{occ+95}
{Oke}, J.~B., {Cohen}, J.~G., {Carr}, M., {Cromer}, J., {Dingizian}, A.,
  {Harris}, F.~H., {Labrecque}, S., {Lucinio}, R., {Schaal}, W., {Epps}, H.  \&
  {Miller}, J. {The Keck Low-Resolution Imaging Spectrometer}.
\newblock {\it Publ. Astr. Soc. Pacific} {\bf 107}, 375--385 April 1995.

\bibitem[{Sivakoff}, {Sarazin} \& {Irwin}<12>]{ssi+03}
{Sivakoff}, G.~R., {Sarazin}, C.~L.  \& {Irwin}, J.~A. {Chandra Observations of
  Low-Mass X-Ray Binaries and Diffuse Gas in the Early-Type Galaxies NGC 4365
  and NGC 4382 (M85)}.
\newblock {\it Astrophys. J.} {\bf 599}, 218--236 December 2003.

\bibitem[{Oke}<13>]{oke81}
{Oke}, J.~B. {A New Cassegrain Spectrograph for the Hale 5-meter Telescope}.
\newblock {\it Bull. American Astron. Soc.} {\bf 13}, 509 March 1981.

\bibitem[{Retter} \& {Marom}<14>]{rm03}
{Retter}, A. \& {Marom}, A. {A model of an expanding giant that swallowed
  planets for the eruption of V838 Monocerotis}.
\newblock {\it Mon. Not. R. astr. Soc.} {\bf 345}, L25--L28 October 2003.

\bibitem[{Arp}<15>]{arp56}
{Arp}, H.~C. {Novae in the Andromeda nebula.}
\newblock {\it Astron. J.} {\bf 61}, 15--34 February 1956.

\bibitem[{Capaccioli} {\it et~al.}<16>]{cdr+89}
{Capaccioli}, M., {della Valle}, M., {Rosino}, L.  \& {D'Onofrio}, M.
  {Properties of the nova population in M31}.
\newblock {\it Astron. J.} {\bf 97}, 1622--1633 June 1989.

\bibitem[{Darnley} {\it et~al.}<17>]{dbk+04}
{Darnley}, M.~J., {Bode}, M.~F., {Kerins}, E., {Newsam}, A.~M., {An}, J.,
  {Baillon}, P., {Novati}, S.~C., {Carr}, B.~J., {Cr{\'e}z{\'e}}, M., {Evans},
  N.~W., {Giraud-H{\'e}raud}, Y., {Gould}, A., {Hewett}, P., {Jetzer}, P.,
  {Kaplan}, J., {Paulin-Henriksson}, S., {Smartt}, S.~J., {Stalin}, C.~S.  \&
  {Tsapras}, Y. {Classical novae from the POINT-AGAPE microlensing survey of
  M31 - I. The nova catalogue}.
\newblock {\it Mon. Not. R. astr. Soc.} {\bf 353}, 571--588 September 2004.

\end{thebibliography}

\bigskip\bigskip

\noindent  {\bf Supplementary Information} is linked to the online version of the paper at www.nature.com/nature.

\bigskip

\noindent  {\bf Acknowledgments:} We thank D. Frail for discussion and constructive criticism.  We
would like to express our gratitude to astronomers who maintain the
NED database at IPAC and the data archives of the Hubble Space
Telescope, the Spitzer Space Telescope and the Chandra X-ray
Telescope.  Our work has been in part by NASA, NSF, the Sylvia and
Jim Katz Foundation and the TABASGO Foundation.

\bigskip

\noindent {\bf Author Information:} Reprints and permissions information is available at npg.nature.com/reprintsandpermissions. The authors declare no  competing financial interests. Correspondence and requests for materials should be adressed to S.R.K (srk@astro.caltech.edu).

\newpage

\newpage

\newpage

\newpage

\end{document}